# A Relaxation-based Network Decomposition Algorithm for Parallel Transient Stability Simulation with Improved Convergence

Jian Shi, Member, IEEE, Brian Sullivan, Member, IEEE, Mike Mazzola, Member, IEEE, Babak Saravi, Student Member, IEEE, Uttam Adhikari, Member, IEEE, and Tomasz Haupt

**Abstract**—Transient stability simulation of a large-scale and interconnected electric power system involves solving a large set of differential algebraic equations (DAEs) at every simulation time-step. With the ever-growing size and complexity of power grids, dynamic simulation becomes more time-consuming and computationally difficult using conventional sequential simulation techniques. To cope with this challenge, this paper aims to develop a fully distributed approach intended for implementation on High Performance Computer (HPC) clusters. A novel, relaxation-based domain decomposition algorithm known as Parallel-General-Norton with Multiple-port Equivalent (PGNME) is proposed as the core technique of a two-stage decomposition approach to divide the overall dynamic simulation problem into a set of subproblems that can be solved concurrently to exploit parallelism and scalability. While the convergence property has traditionally been a concern for relaxation-based decomposition, an estimation mechanism based on multiple-port network equivalent is adopted as the preconditioner to enhance the convergence of the proposed algorithm. The proposed algorithm is illustrated using rigorous mathematics and validated both in terms of speed-up and capability. Moreover, a complexity analysis is performed to support the observation that PGNME scales well when the size of the subproblems are sufficiently large.

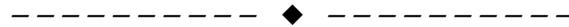

## 1 INTRODUCTION

DYNAMIC simulation has always been an essential tool to study and evaluate power systems [1], [2]. As design, planning, analysis and operation of a large and complex interconnected power system would inevitably involve the evaluation, understanding and prediction of the dynamic behaviors of the system under a wide range of scenarios, computational simulation provides an effective and essential tool to facilitate this requirement throughout the whole design process.

One of the major challenges for power system dynamic simulation is on the scale and complexity of the current and future power grid. As the backbone of the power system, electric grids all over the world are undergoing fundamental revolutions. Take the US power grid for example, according to the U.S. Department of Energy, a national electricity backbone will be built to link the east and west coasts, as well as Canada and Mexico by year 2030 to give customers "continental" access to energy supplies [3], [4]. As part of this envisioned national grid, according to Oak Ridge National Lab's estimation, the Eastern Interconnection model (EI) at year 2030 will be expanded to include over 70,000 buses and 8,000 generators, compared with the current EI model which only contains 16,000 buses and 3,000 generators. It is evident that the power system of the future will be a much larger interconnected system on a scale that has never been encountered before. This, combined with the wide deployment of smart-grid technologies, such as mini- and micro-grids and distributed energy technologies, has added significant complexity to the already sophisticated structure of the interconnected power grid. There is a growing realization that contribution from the computer industry will directly affect and influence the shape of the next-generation power grid.

However, as pointed out in [3-5], the application of advanced computing techniques in power and energy industry has significantly lagged behind other industries. Legacy codes and algorithms written back in the 80's based on single-process and serial operation are still dominating in the current power system analysis tools. To fully cope with the scale and complexity of the problem and adequately capture the sophisticated dynamic interactions and interdependencies, new power system simulation techniques need to be developed to accommodate the accelerated growth of the size, complexity and heterogeneity of the problem and provide improved computational efficiency, accuracy, capacity, and scalability. Meeting this challenge requires both the introduction of parallelism in algorithm design and the advanced computing platforms such as High Performance Computing (HPC) cluster [5-7].

While the multi-core, parallel computing hardware provide the necessary computational capabilities, parallel processing adds the dimension of concurrency and brings benefits such as speed-up, capability and scalability. To

————————————————

- *J. Shi, B. Sullivan, M. Mazzola, B. Saravi and T. Haupt are with the High Performance Computing Collabotory at Mississippi State University, Starkviile, MS 39759 E-mail: Jian@dasi.msstate.edu; Sullivan@cavs.msstate.edu; Mazzola@ece.msstate.edu; Saravi@dasi.msstate.edu*
- *U. Adhikari is with PEAK Reliability, Loveland, CO 80537 E-mail: uttamadhikari@gmail.com*

achieve parallelism, a natural approach to undertake is through Domain Decomposition Methods (DDM) [8]. In the context of power system dynamic simulation, DDM generally refers to a class of techniques that decompose the original problem defined over a domain into smaller "*subproblems*" on overlapping or non-overlapping subdomains and coordinate the solution among subproblems to yield the equivalent solution of the original problem.

This manuscript proposes a novel relaxation-based decomposition technique named Parallel-General-Norton with Multiple-port Equivalent (PGNME) to facilitate the distributed dynamic simulation and analysis for large-scale and very large-scale power systems on HPC platforms. The design algorithm of PGNME can be seen as a combination of two techniques: a Jacobi-like Parallel Updating Relaxation (PUR) process and a novel ME based Spectral Radius Reduction (SRR) technique. While the PUR algorithm is introduced to reconcile the solutions of subproblems until reaching a global convergence, in this paper, we particularly focus on demonstrating that through the analytical modeling of the convergence properties of PUR, an effective and scalable SRR method can be derived as the preconditioning mechanism to tune the configurations of PUR in order to achieve significantly enhanced global convergence. We will start by illustrating the formulation of PGNME method for a general multiple-subgraph, multiple-port system, followed by demonstrating PGNME on an intuitive single-port system containing two-subgraphs and a more complicated example system with 3 subgraphs and a total of 6 ports.

The main contribution of this paper can be summarized as the following aspects:

- **System size and degree of parallelism**: While previous work [6-7], [9], [11-27] (please refer to Section 2.1 for a comprehensive literature review) have been successful in achieving parallelism for power system dynamic simulation, they are designed for and validated based on smaller scale problems, and thus generally limited to relatively lower degree of concurrency in the range of 4- to 20-way parallelism for system size in the order of 10000 buses. In this manuscript, we realize that the electric grid is undergoing a fundamental change. The size and complexity of the future power grid is envisioned to reach a scale not witnessed since its creation and far beyond the scopes of existing studies presented in the literature. Therefore, our research effort was dedicated to exploring the capabilities and limitations of simulating truly large-scale and ultra-large-scale power systems on the scale of one million buses. Although power transmission systems of this scale do not exist yet, when the distribution system and portions of the load "beyond the meter" are included, the scale is at least this size. The research effort reported here is valuable for understanding and evaluating the bottlenecks of the DDMs when the size and complexity of the problem become critical.
- **Improved convergence**: In the context of large-scale power system dynamic simulation, the existing relaxation-based decomposition methods have been criticized for the sensitivity to partitioning schemes and lack of control over convergence especially when the couplings between subproblems are strong. Most of the literature addresses this issue by solely relying on the utilization of effective/customized graph partitioning techniques [16], [23] [28-29]. However, the feasibility and effectiveness of this approach when applied to a large-scale problem is in doubt. In this paper, a novel design feature of the proposed algorithm is to formulate the iterative parallel solution process of the power system network into a rigorously defined analytical model. Then by exploiting the structural properties of this model, the convergence of the process can be controlled. Simulation results indicate that this feature effectively addresses the issue of slow and uncontrollable convergence.
- **Hardware implementation**: Most existing parallel algorithms and simulation routines developed in the state-of-the-art power system simulator are targeting shared-memory parallel computing architectures and small/medium size distributed-memory clusters. Instead, in this manuscript, realizing the scale and complexity of emerging problems, the proposed algorithm aims to leverage the massive computational capacity offered by the latest HPC infrastructures and software programming models. By developing, testing and validating the performance of the proposed algorithm on a 110-node, 2200- modern HPC cluster, it is demonstrated for the first time that massive power system dynamic simulation is feasible. In [30], it is also demonstrated that faster-than-real-time performance can be obtained using the proposed algorithm with the smallest time step on the largest system model as compared to existing literature using similar modeling strategies.
- **Scalability**: A thorough complexity analysis has been performed to assure the performance of PGNME to solve a truly large-scale problem on a large machine configuration. It has been demonstrated analytically that the proposed algorithm scales well with the number of processors and the size of the system under study.

The paper is organized as follows: in Section 2, general backgrounds of the research work presented in this paper is briefly reviewed. In Section 3, the general derivations of PGNME algorithm are explained. In Section 4, PGNME is applied to two test systems to illustrate the detailed formulation of PGNME. In Section 5, the implementation of PGNME on HPC cluster is presented while the simulation results obtained from two case studies are reported. The conclusion of this paper is drawn in Section 6.

## 2 BACKGROUND

Power system dynamic phenomena are highly complicated and can range from microseconds to days, therefore different categories of modeling and simulation techniques exist to efficiently and adequately capture the system behaviors over different frequency ranges [1]. In this paper, we limit our scope to the category of Transient Stability (TS) simulation which focuses on evaluating system response to large and sudden disturbances to assure secure and stable operation of the power system. As the goal of

TS simulation is to effectively capture the low frequency electromechanical dynamic phenomenon assuming single-phase fundamental frequency behavior of the transmission network, it is arguably the most suitable approach to study a large, interconnected power network. To set the stage for the rest of the paper, the general background of large-scale TS simulation, the most-recent DDM techniques that have been adopted in parallel dynamic TS simulation, as well as the two-level decomposition hierarchy adopted in this work is briefly overviewed in this section.

### 2.1 Related work

In the context of parallel power system TS simulation, there are generally two separate approaches existing in DDM [6], [7], [9], [11-27]: fine-grained and coarse-grained. A summary of the most-recent DDM for power system TS simulation can be found in Table. 1. However, Table 1 should not be construed as exhaustive of all applicable literature, and specifically does not intend to preclude the use of other domain decomposition methods well known to the field of computational fluid dynamics (CFD) in the area of unstructured grids, which may be attractive.

In the fine-grained approach [6], [7], [11-14], the parallelism is mainly obtained through matrix/vector reduction techniques to exploit the structural properties (such as sparsity and repetition) of the linear power system network matrices and make it suitable for parallel implementation. On the other hand, the coarse-grained approach [9], [16-18], [20-27] is applied directly to the system of equations. Compared with fine-grained methods, coarse-grained methods have manifested various advantages, such as: the parallelism is not limited by the block structure of the system, localized information for the subproblems is kept, subproblems can be solved using locally adapted simulation settings, etc. [10]. While coarse-grained parallelism can be further exploited from different perspectives such as spatial parallelism [9], [17-18], [23-24], time parallelism and spatial-time parallelism (e.g. waveform relaxation) [16], [20-21], [25-27], in this paper, we mainly focus on the spatial parallelism (parallel-in-space) which incorporates a partitioning scheme to first decouple the original problem into various pieces then solve each of the partitioned subproblems in parallel at a given time instant.

The construction of a DD algorithm involves essentially three steps [8-10]. First of all, an appropriate partition scheme is adopted to divide the original problem into subproblems that can be solved concurrently. Secondly, the interdependencies between each subproblem need to be specified, formulated in the form of "*interface variables*", and solved along with the "*interior variables*" which are coupled only through local equations of a subproblem. In the third step, all the subproblems need to be reconciled through the interface variables to derive a coordinated, converged global solution. Based on the way the interface variables are computed, spatial DD techniques can be classified into two main approaches: Schur-complement method [9], [17], [18], [24] and Schwartz alternating method [23].

In Schur-complement method, a reduced system model needs to be formulated to specifically calculate the interface variables. The subproblems then become naturally decoupled. Therefore, it is considered a "direct" method.

On the other hand, Schwarz alternating method "relaxes" the interface variables during the subproblem solution [10], [23]. Once the associated subproblems are solved, the interface variables are updated and this process is iterated until a global convergence is achieved. Schwarz alternating method is more attractive due to the fact that no additional reduced system needs to be formulated and calculated, thus the overall solution is completely parallelizable. The main drawback of Schwarz alternating method is the convergence properties as excessive number of iterations may diminish its merits. However, in the following discussion, we will focus on demonstrating that while PGNME is developed based on the concept of Schwarz alternating method, the convergence enhancement featured in PGNME preconditioning allows it to fully exploit the advantages of Schwarz domain decomposition.

TABLE 1. A summary of the most-recent representative DDMs for power system TS simulation

| Author | Year | Network Size (# of buses) | DD Methodology | Targeted Platform |
|---|---|---|---|---|
| Jalili-Marandi *et al*. [14][15] [20][21] | 2009[20][21], 2010[14], 2012[15], | 39 [20][21][14], 9984 [15] | Fine-Grained LU [14] Fine-Grained LU + Coarse-Grained IR [15] Coarse-Grained Instantaneous Relaxation (IR) [20][21] | Distributed memory PC [20][21] GPGPU [14][15] |
| Huang *et al*. [11][12][13] | 2013[11], 2017[12], 2017[13] | 16072 [11], 17000 [12], 16072 [13] | Fine-Grained Woodbury/LU [11] Fine-Grained LU [12][13] | Shared memory HPC [11] Distributed memory HPC [12] Distributed/Shared memory HPC [13] |
| Flueck *et al*. [22][23] | 2012[22], 2012[23] | 7935 [22], 2383 [23] | Fine-Grained VDHN/LU [22] Coarse-Grained BJP + GMRES [23] | Distributed Memory HPC [22] Shared Memory HPC [23] |
| Aristdou *et al*. [9][24] | 2014[9], 2016[24] | 2565 [24] 15226 [9] | Coarse-Grained Schur-complement [9][24] | Share memory laptop/HPC [9][24] |
| Marti *et al*. [17][18] | 2009[17], 2013[18] | 15000 [17], 14327 [18] | Coarse-Grained Schur-complement [17][18] | Dedicated PCI cards and shared memory hub [17][18] |
| Pruvost *et al*. [25] | 2011 | 15350 | Coarse-Grained WR/Epsilon decomposition | Shared memory HPC |
| Liu *et al*. [16] | 2016 | 12685 | Coarse-Grained WR/Adomian decomposition | Shared memory HPC |
| **Shi *et al*.** | **2017** | **258066** | **Coarse-Grained Schwartz/PGNME preconditioning** | **Distributed memory HPC** |

## 2.2 Problem formulation

In the scope of TS simulation, the simulation of power system dynamics is performed by solving a large differential-algebraic initial value problem in the time-domain that can be represented by a set of first-order differential equations [1] in the form of:

$$\dot{\mathbf{x}} = \mathbf{f}(\mathbf{x}, \mathbf{y}, t), \mathbf{f} : \mathbb{R}^{n+m} \to \mathbb{R}^n \quad (1\text{-a})$$
$$\mathbf{x}(t_0) = \mathbf{x_0}, \mathbf{x_0} \in \mathbb{R}^n$$

and an algebraic equation set of:

$$0 = \mathbf{g}(\mathbf{x}, \mathbf{y}, t); \mathbf{g} : \mathbb{R}^{n+m} \to \mathbb{R}^m \quad (1\text{-b})$$

where $\mathbf{x} \in \mathbb{R}^n$ denotes the vector of differential state variables, $\mathbf{y} \in \mathbb{R}^m$ denotes the vector of algebraic state variables, $t$ denotes the time instant, and $\mathbf{x_0}$ denotes the values of the vector $\mathbf{x}$ at the initial time instant $t_0$.

## 2.3 A two-stage decomposition structure

Under the common assumption that every component in the power system is coupled to the other components only through the network and the network is in the sinusoidal steady state within the bandwidth of interest, Eq. (1-a) and Eq. (1-b) can be further divided and rearranged into two sub-sets:

$$\dot{\mathbf{x}}_C = \mathbf{f}_C(\mathbf{x}_C, \mathbf{y}_C, \mathbf{y}_N, t) \quad (2\text{-a})$$
$$0 = \mathbf{g}_C(\mathbf{x}_C, \mathbf{y}_C, \mathbf{y}_N, t) \quad (2\text{-b})$$

and

$$0 = \mathbf{g}_N(\mathbf{y}_C, \mathbf{y}_N, t) \quad (3)$$

Once these two sub-sets of equations are constructed, the Alternating Solution Method (ASM) [2] which solves Eq. (2) and Eq. (3) alternately can be utilized to facilitate the proper separation and parallel simulation of dynamic components and the linear network. Sub-set Eq. (2) is a collection of independent, decoupled sub-sets corresponding to each generator and load. It includes all the differential and algebraic equations that describe the synchronous machines, their control devices such as excitation controls and turbine governors, and load dynamics. To solve Eq. (2), the sub-set of differential equations as defined in Eq. 2 (a) is firstly discretized using a specific numerical integration method. Then, the resulted algebraic equations are combined with the sub-set of algebraic equations as defined in Eq. 2(b). This combined set of algebraic equations (usually non-linear) can then be solved using a specific non-linear algebraic solver. The solution to this stage is in the form of current injections at the buses where the dynamic components are attached. Therefore, once Eq. (2) is solved, the vector of current injection at each bus, namely, $\mathbf{I}$, can be determined. This is considered the first stage of decomposition.

Following stage 1, the vector of current injection $\mathbf{I}$ will be fed into sub-set Eq. (3), which is the nodal network equation that can be represented using linear, sparse admittance matrix equations in the form of $\mathbf{V} = \mathbf{Y}^{-1}\mathbf{I}$ where $\mathbf{Y}$ represents the $n \times n$ nodal admittance matrix, $\mathbf{V}$ denotes the vector of voltages at each bus. Essentially, the purpose of this stage is to solve a large-scale linear and sparse matrix equation.

Using this decomposition strategy, sub-set Eq. (2) is completely separable and the overall system is coupled only through sub-set Eq. (3). It is apparent that stage 1 is component-wise independent and easily parallelizable (each dynamic component has their own individual subset of Eq. 2(a) and Eq. 2(b)). The most challenging problem for decomposing the solution of transient simulation when the scale of the electrical network is very large lies in the proper decomposition of the linear passive network Eq. 3 that needs to be solved in Stage 2. Therefore, in this paper, we focus on addressing this issue by developing a self-consistent parallel simulation solution known as PGNME to tackle the most significant obstacle and bottleneck in parallelizing the dynamic simulations by decomposing Eq. (3) into subgraphs that can be solved together with their associated dynamic components in parallel. Although PGNME is intended to work as a part of the self-consistent solution of all equations as shown in Fig. 1 (a), in this manuscript our focus is on deriving and demonstrating PGNME as a linear network decomposition technique for solving the computationally intensive part of the problem, which are the power network equations. An application of PGNME as part of a full transient stability simulation is published in [30]. Therefore, for the rest of the discussion, a simplified computation paradigm shown in Fig. 1 (b) is adopted where the computation of stage 1 was replaced by pre-computed data that simulates the solution of Eq. (2-a) and Eq. (2-b). However, in [30], the complete simulation routine shown in Fig. 1(a) was tested and very satisfactory results were obtained.

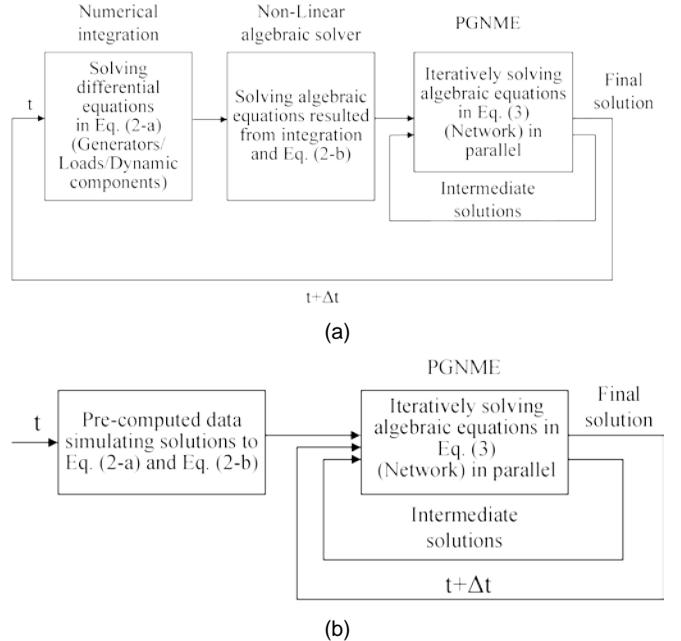

Fig. 1. A standard ASM solution procedure is shown in (a) while the simplified alternating solution considered in this manuscript is shown in (b)

## 2.4 Solving the Algebraic Equations by Iterative Relaxation

Generally, a linear (either real or complex) system can be represented in the form of:

$$\mathbf{AX} = \mathbf{b} \qquad (4)$$

where **A** is nonsingular and **b** is a constant vector. If **A** is nonsingular, then Eq. (4) can be iteratively solved with the linear stationary method using a parallel updating strategy [10]. Fig. 2 illustrates a graph-based decomposition of a system represented in Eq. (4). The graph consists of vertices where state-variables of the potential type are defined, connected by edges where state-variables of the flux type are defined. The solution to Eq. (4) represents a self-consistent balance between the fluxes on the edges and the potentials at the vertices that cause flux to circulate. The parallel decomposition into subproblems involves cutting select edges in the larger graph in order to form subgraphs isolated from all other subgraphs. The problem of the solver is to reconcile cut-edge flux linkages to be equal as they would be in the original graph by using the subproblem solvers to compute the reduced order equations in the form of Eq. (4). This of course requires specification of boundary conditions at all cut edges to close the set of equations. The method of the "*boundary bus*" will be described in the next section to accomplish this. For now, it is adequate to recognize the bifurcation of the edge flux variables into $2n$ in number, where $n$ is the number of edges cut. A numbering scheme is suggested in Fig. 2 where at the $j$-th cut edge, one of these variables is $X_{2j-1}$, while the other is $X_{2j}$. The objective of the solver is to relax the difference between all pairs of these cut edge flux variables through a finite number of iterations until all $2n$ variables converge, i.e., such that for all $j$, $X_{2j-1} = X_{2j}$ within the tolerance for convergence. Since the set of equations for each subgraph are solved at every iteration count $i$, then as the cut-edge variables converge all the state variables converge with them.

Forming all cut-edge variables into a column vector **X** at every relaxation iteration $l$ leads to an iteration equation of the first degree in the form of:

$$\mathbf{X}^{l+1} = \mathbf{W}\mathbf{X}^{l} + \mathbf{G} \qquad (5)$$

where **X** denotes the vector of cut-edge state variables that are updated iteratively, **G** is a constant vector, and **W** denotes the iteration matrix. **X** is a column vector of length $2n$, **W** is a $2n \times 2n$ square matrix. Examples of deriving **W** and exploiting its structure to enhance the convergence of Eq. (5) will be covered in the following section.

It is obvious from the state flow diagram shown in Fig. 2 that a cut-edge state variable does not update itself until the second iteration interval is complete. Thus, **W** has the peculiar property of zeros on the main diagonal, which is a feature of the parallel procedure used.

The iteration matrix **W** directly affects the iterative process. Theorem 4.1 from [10] indicates that if **W** is a square matrix such that the spectral radius $\rho(\mathbf{W}) < 1$, then the iteration process Eq. (5) converges for any **G** and $\mathbf{X}^0$. Meaning that for any iterative function and any initial condition the process will converge if $\rho(\mathbf{W}) < 1$. Within this definition, $\rho(\mathbf{W})$ is a scalar regardless if **W** is real or complex in the form of:

$$\rho(\mathbf{W}) = \max(|\lambda_1|, |\lambda_2|, \ldots |\lambda_n|) \qquad (6)$$

where $\lambda_1, \ldots \lambda_n$ denotes the eigenvalues of the matrix **W**. More specifically, Equation (2-2.8) and (2-2.10) from [10] have defined that if $\rho(\mathbf{W}) < 1$, then the average rate of convergence can be approximated [10] in the form of:

$$R_\infty(\mathbf{W}) = -\log(\rho(\mathbf{W})) \qquad (7)$$

This indicates that the closer $\rho(\mathbf{W})$ gets to zero, the faster convergence can be achieved. Especially when $\rho(\mathbf{W})$ approaches 0, the convergence rate becomes infinite which indicates an ideally "instantaneous" convergence (i.e. 1 iteration).

It will be shown in section 4.1 that for a two-subgraph partition, by formulating the explicit analytical derivation of **W** and performing a "boundary bus" preconditioning step, the spectral radius can be set ideally to zero. This process is named Parallel General Norton (PGN). However, PGN itself is not scalable beyond two partitions, making this preconditioning step unusable for large-scale high-performance cluster computing. The purpose of this paper is to introduce an important extension of PGN: PGNME, which will greatly extend the method of [35] and [36] to an arbitrary number of partitions. The demonstration of the scalability of PGNME will include the conditions under which small spectral radius, and thus minimal relaxation iterations and maximum parallel computing gain, can be forced by a set of preconditioning tests on an arbitrary partition.

In the following section, it is shown that a decomposed linear power system network can be solved in parallel with rapid convergence by manipulating the spectral radius $\rho(\mathbf{W})$ through proper selection of boundary admittance values by the PGNME process.

## 3 ALGORITHM DEVELOPMENT

In this section, the derivation of the complete algorithm of PGNME is demonstrated.

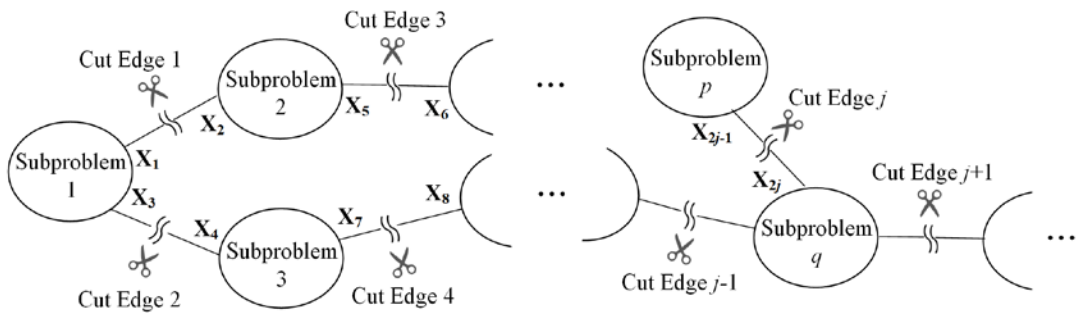

Fig. 2. Illustration of the decomposition of the graph representing (12) into subgraphs formed by cutting $n$ edges in the original graph. At every $j$th cut edge two variables are formed, $X_{2j-1}$ and $X_{2j}$, where only one edge state-variable existed before.

## 3.1 Boundary bus definition

To illustrate the PGNME algorithm, we will start by explaining the graph representation of the power system network and how the interface variables are defined in this paper to decouple a network.

As a geographically distributed system, a power system network can be conveniently represented as an interconnected graph. Each bus is treated as a node, and each branch is treated as an edge. Based on a given graph partitioning algorithm, appropriate edge cuts can be made, and then the original graph is decomposed into various subgraphs. Throughout this paper, these subgraphs are referred as "*subsystems*". Then the cut edges between the subsystems can be represented by units named "*boundary buses*". Boundary buses act as the interface to complete each subsystem and represent the effects of external system states on the boundary conditions of a specific subsystem. Since boundary buses are introduced to represent linear subgraphs, they can be simply represented by their Norton equivalent forms, consisting of a controlled current source, namely, a *boundary state variable* (denoted by $S$) in parallel with a *boundary admittance* (denoted by $G$). It also can be defined that each boundary bus is connected through a "*port*" to its attached subsystem. After terminating each cut edge with a boundary bus, a complete "*subproblem*" can be obtained. A subproblem can also be referred to as a partition. It is expected that boundary buses can reconcile with their associated subgraphs in a coordinated way to ensure the interface conditions between subproblems are satisfied, and thus to guarantee the validity of the solution.

As an example, as shown in Fig. 3 (a), we consider a general system that is decomposed into various subsystems. It is assumed that within the many subsystems, subsystem $X$ is connected to subsystem $Y$ through an edge $y_{XY}$ which now becomes a cut edge. If we denote the ports resulting from cut edge $y_{XY}$ port $j$ ($j \in [1,n]$) for subsystem $X$ and port $k$ ($k \in [1,m]$) for subsystem $Y$, the representation of the boundary bus $j$ and $k$ is shown in Fig. 3 (b). Assume that subsystem $X$ has a total of $n$ edges connected to it that are cut and subsystem $Y$ has a total of $m$ edges connected to it that are cut. Based on the previous boundary bus definition, subsystem $X$ can be considered an $n$-port system and similarly subsystem $Y$ can be considered an $m$-port system.

## 3.2 PUR method

In this subsection, the general concept of the Parallel Updating Relaxation (PUR) strategy is explained.

Without losing generality, consider subsystem $X$ as an example. Since each subgraph resulting from the graph-based partitioning can be considered a linear passive network, simulating subsystem $X$ requires solving the following network equation:

$$\mathbf{V}_{\text{sub}} = \mathbf{Y}_{\text{sub}}^{-1} \mathbf{I}_{\text{sub}} \quad (8)$$

where $\mathbf{I}_{\text{sub}}$ and $\mathbf{V}_{\text{sub}}$ denote the node current and voltage within subsystem $X$ respectively, $\mathbf{Y}_{\text{sub}}$ denotes the nodal admittance matrix. After the boundary buses are added, under the assumption that the boundary bus voltage is denoted as $V_{dxj}$ ($j \in [1,n]$) and the boundary bus current injection is denoted as $S_{xj}$, then the expanded nodal admittance matrix can be defined in the form of:

$$\mathbf{V}_{\text{mod}} = \mathbf{Y}_{\text{mod}}^{-1} \mathbf{I}_{\text{mod}} \quad (9)$$

where $\mathbf{V}_{\text{mod}} = [\mathbf{V}_{\text{sub}}^{\text{T}}\ V_{dx1}\ V_{dx2}\ \ldots\ V_{dxn}]^{\text{T}}$, $\mathbf{I}_{\text{mod}} = [\mathbf{I}_{\text{sub}}^{\text{T}}\ S_{x1}\ S_{x2}\ \ldots\ S_{xn}]^{\text{T}}$, and the expanded admittance matrix with boundary buses included is denoted as $\mathbf{Y}_{\text{mod}}$.

Particularly, at the port $j$ of subproblem $X$, based on the fundamental Kirchhoff's Current and Voltage Laws (KCL and KVL), a set of algebraic equations governing the interactions of subsystem $X$ and boundary bus $j$ can be derived in the form of:

$$V_{dxj} - V_{xj} - I_{xj} y_{xy}^{-1} = 0 \quad (10\text{-a})$$

$$G_{xj} V_{dxj} = S_{xj} - I_{xj} \quad (10\text{-b})$$

As the equivalent representation of subproblem $Y$ is symmetric to subproblem $X$, the same form of equations as used in Eq. (10) work for subproblem $Y$ as well. In order for boundary bus $k$/boundary bus $j$ to reproduce the subsystem $Y$/subsystem $X$ accordingly, the constraints that link subproblem $X$ to subproblem $Y$ include the current flowing through the admittance of the cut branch $y_{xy}$ and the terminal voltage at the ports $V_{xj}$ and $V_{yk}$. Based on the aforementioned discussion on iterative methods, these two constraints can be relaxed and expressed in the Jacobi-like iterative form of:

Subproblem $X$: $\begin{cases} V_{dxj}(l+1) = V_{yk}(l) \\ I_{xj}(l+1) = -I_{yk}(l) \end{cases}$ (11-a)

Subproblem $Y$: $\begin{cases} V_{dyk}(l+1) = V_{xj}(l) \\ I_{yk}(l+1) = -I_{xj}(l) \end{cases}$ (11-b)

where $l$ is the iteration index. Based on Eq. (11), the corresponding parallel updating strategy for the controllable boundary state variable $S_{xj}$ and $S_{yk}$ can be obtained as:

$$S_{xj}(l+1) = G_{xj} V_{dxj}(l+1) + I_{xj}(l+1) = G_{xj} V_{yk}(l) - I_{yk}(l) \quad (12\text{-a})$$

$$S_{yk}(l+1) = G_{yk} V_{dyk}(l+1) + I_{yk}(l+1) = G_{yk} V_{xj}(l) - I_{xj}(l) \quad (12\text{-b})$$

This updating routine is repeated every relaxation iteration until the convergence criteria is met. In this case, the mismatch between the port currents is chosen as the convergence criteria in the form of:

$$\left\| I_{xj}(l) + I_{yk}(l) \right\| \leq \sigma \quad (13)$$

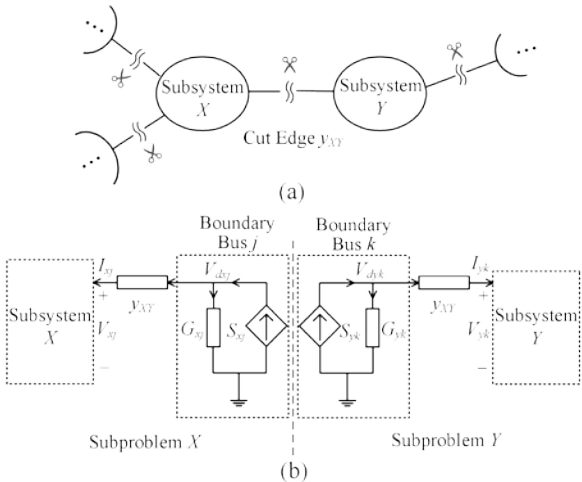

Fig. 3. Boundary bus creation and reconciliation

where σ is the convergence threshold. For all simulation results shown in section 5, σ =1×10⁻⁶ is used.

Therefore, beginning in the *l*-th relaxation iteration, the reconciliation process for each subproblem using the PUR technique can be summarized as follows:

Step.1: Apply $S_{x1}(l)$, $S_{x2}(l)$… $S_{xn}(l)$ at each boundary bus of subproblem *X* and $S_{y1}(l)$, $S_{y2}(l)$,… $S_{ym}(l)$ at each boundary bus of subproblem *Y* respectively.

Step.2: Solve Eq. (9) to obtain the boundary bus voltages $V_{dx1}(l)$, $V_{dx2}(l)$, …. $V_{dxn}(l)$ and all the port voltages $V_{xj}(l)$ (*j*=1,2…*n*) and then Eq. (11-a) for currents $I_{xj}(l)$ for subproblem *X*. The same procedure can be performed on subproblem *Y* to derive the port voltage $V_{yk}(l)$ and current $I_{yk}(l)$ (*k*=1,2…*m*) for each port using Eq. (11-b).

Step.3: Update $S_{xj}(l+1)$ (*j*=1,2…*n*) based on Eq. (12-a) for subproblem *X* and update $S_{yk}(l+1)$ (*k*=1,2…*m*) based on Eq. (12-b) for subproblem *Y*.

Step.4: Check convergence criteria described by Eq. (13) for each port between subsystem *X* and *Y* as well as the iteration number limit, if the constraints for all the ports are satisfied, then the parallel updating process ends; if not satisfied, *l*=*l*+1 and PUR repeats at Step 1.

Note that while the PUR steps are defined and explained in the context of subproblem *X* and *Y*, the same reconciliation procedure applies to all the boundary buses across the ensemble of subproblems. At each relaxation iteration, all of the subsystems are solved and then reconciled in a simultaneous manner until a global convergence, which takes account of the boundary conditions of all the subproblems, is satisfied.

Now we have developed the reconciliation strategy for the basic PUR method. However, as mentioned in the previous section, a critical issue that affects the performance of this process is the convergence rate.

A common way that has been suggested in the literature lies in the network partitioning algorithm. It is recommended that by exploiting the inherent topological structure and unique properties of the power grid (such as stiffness, coherency of machines, localized phenomena), subproblems that have strong self-similarities and weak interconnections to other subproblems can be effectively identified. This approach lacks three critical aspects: 1) the support of rigorous mathematics to understand and analyze the underlying relationship between system characteristics and convergence properties to design an "optimal" iterative process; 2) robustness when the complexity of the problem grows with the size of the network; 3) when global partitioning across ownership boundaries is not an available option due to reasons such as privacy and security. In the following discussion, we shall focus on demonstrating a spectral radius reduction mechanism that exploits the analytical convergence properties of the proposed PUR method and the aforementioned convergence evaluation techniques to achieve improved convergence through preconditioning of the sub-problems rather than relying solely on the partitioning strategy. Later, it will be concluded that preconditioning can be augmented by partitioning optimization to maximize the gain from parallel computing.

### 3.3 PGNME preconditioning

Since each subsystem is linear, according to the multi-port equivalent theory, an equivalent form can be computed for each subsystem in terms of each port. Consider an arbitrary subsystem *X*, the multi-port equivalent of which can be defined in the following form:

$$\begin{bmatrix} I_{x1} \\ I_{x2} \\ \vdots \\ I_{xn} \end{bmatrix} = \begin{bmatrix} y_{x1} & h_{x21} & \cdots & h_{xn1} \\ h_{x12} & \ddots & & h_{xn2} \\ \vdots & & & \vdots \\ h_{x1n} & h_{x2n} & \cdots & y_{xn} \end{bmatrix} \begin{bmatrix} V_{x1} \\ V_{x2} \\ \vdots \\ V_{xn} \end{bmatrix} + \begin{bmatrix} D_{x1} \\ D_{x2} \\ \vdots \\ D_{xn} \end{bmatrix} \quad (14)$$

where $I_{xj}$ denotes the current flowing into port *j* ( $j \in [1,...n]$ ), $V_{xj}$ denotes the voltage at port *j*, $y_{xj}$ denotes the self-admittance of port *j*, $h_{xji}$ ( $i \in [1,...n], i \neq j$ ) indicates the port coupling parameters between port *i* and port *j* within the subsystem *X*, more specifically, between port current $I_{xj}$ at port *j* and port voltage $V_{xi}$ at port *i*, and $D_{xj}$ denotes the effects of active current sources within subsystem *X* at port *j*. Similarly, other subproblems can be described using their multi-port equivalent representations in the form of Eq. (14) as well.

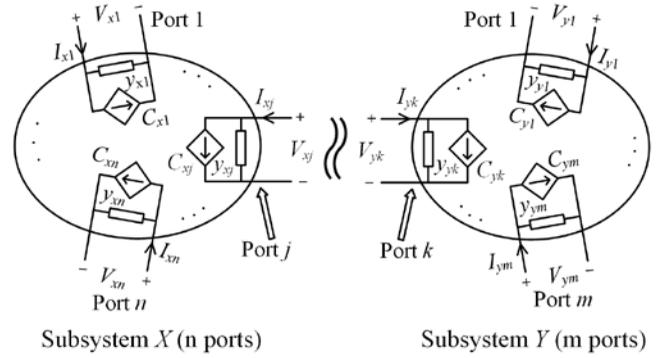

Fig. 4. The multi-port equivalent representation of subsystem X and Y

#### 3.3.1 Deriving the row entries of the W matrix

Eq. (14) suggests that for port *j* ( $j \in [1,...n]$ ), the following representation is valid:

$$I_{xj} = y_{xj}V_{xj} + C_{xj} \quad (15)$$

where $C_{xj} = \sum_{i=1, i \neq j}^{n} V_{xi}h_{xji} + D_{xj}$. This indicates that as seen at port *j*, subsystem *X* can be represented using an equivalent controlled current source $C_{xj}$ and an equivalent admittance $y_{xj}$ as shown in Fig. 4. $C_{xj}$ accounts for the effects of voltages at the other ports and the active current sources within the subsystem. The value of $C_{xj}$ is directly determined by the off-diagonal elements of the multi-port equivalent matrix $h_{xji}$, the current voltages at the other ports, and a constant $D_{xj}$. Meanwhile, the self-dependency between the port voltage and port current at port *j* is represented using the diagonal element $y_{xj}$. The multi-port equivalent matrix can be obtained through the following steps:

1. Set all port voltages to zero, compute the current at each port to determine $D_x$ for each port.

2. Set the port voltage to zero at all ports but port *j*;

3. Apply an arbitrary test voltage $V_{test}$ ($V_{test}$ = 1 V) at port *j*, solve the subsystem, and compute the current at each port. Then the diagonal entries and the off-diagonal entries

of the multi-port equivalent matrix can be determined. This process can be described as:

$$y_{xj} = \left( \left. (I_{xj} - D_{xj}) \right|_{V_{xj}=V_{test} \text{ and } V_{xi(i \neq j)}=0} \right) / V_{test} \quad (16\text{-a})$$

$$h_{xji} = \left( \left. (I_{xi} - D_{xi}) \right|_{V_{xj}=V_{test} \text{ and } V_{xi(i \neq j)}=0} \right) / V_{test} \quad (16\text{-b})$$

Therefore, by solving subsystem $X$ using Eq. (8) $n$ times with the specified boundary conditions, the complete multi-port equivalent formulation as shown in Eq. (14) can be derived.

Similarly, the same form of multi-port equivalent representation can be derived for subsystem $Y$. Combining the general representation of the boundary bus and the equivalent representation of the subsystem, at port $j$ and port $k$, the equivalent schematic can be derived as shown in Fig. 5. Instead of using Eq. (9) to calculate the boundary bus voltage (which will continue to be used in the actual simulation following the application of the PGNME preconditioning step) the following KCL equation can be derived at port $j$ using the multi-port equivalent representation:

$$-C_{xj} + I_{xj} - y_{xj}V_{xj} = 0 \quad (17)$$

where the $j$th row of Eq. (14) (excluding the diagonal entry) forms the function $C_{xj}$.

Substitute Eq. (10-b) into Eq. (10-a) to eliminate $V_{dxj}$ and write it in the iterative form:

$$\left( y_{xy}^{-1} + G_{xj}^{-1} \right) I_{xj}(l+1) = G_{xj}^{-1} S_{xj}(l+1) - V_{xj}(l+1) \quad (18)$$

Now, substitute Eq. (12-a) into Eq. (18) to eliminate $S_{xj}(l+1)$,

$$V_{xj}(l+1) + \left( y_{xy}^{-1} + G_{xj}^{-1} \right) I_{xj}(l+1) = V_{yk}(l) - G_{xj}^{-1} I_{yk}(l) \quad (19)$$

This equation when repeated to form closed systems of equations that are solved for each boundary bus current is the basis for each row of the iteration matrix **W**.

To eliminate $V_{xj}$ and $V_{yk}$ from both sides of Eq. (19), it is recognized that Eq. (14) can be solved for each port voltage on either side of the boundary bus and substituted into Eq. (19). Applying Cramer's rule to this system allows $V_{xj}(l+1)$ to be solved as:

$$V_{xj}(l+1) = \frac{\det(\mathbf{A}_j)}{\det(\mathbf{A})} \quad (20)$$

where **A** refers to the multi-port equivalent matrix defined in Eq. (14).

Applying the Laplace expansion on the $j$th column to the numerator of Eq. (20) expands it to

$$V_{xj}(l+1) = \frac{I_{xj}(l+1)(-1)^{2j} \det(\mathbf{M}_{jj})}{\det(\mathbf{A})} + \sum_{i=1, i \neq j}^{n} I_{xi}(l+1) a_i + K_x \quad (21)$$

where $a_i = \dfrac{(-1)^{i+j} \det(\mathbf{M}_{ij})}{\det(\mathbf{A})}$ and $K_x$ is a constant resulting from the active current sources $D_x$.

For subsystem $Y$, a similar procedure is applied to solve for $V_{yk}(l)$ by performing the Laplace expansion on the $k$th column. This results in

$$V_{yk}(l) = \frac{I_{yk}(l)(-1)^{2k} \det(\mathbf{M}_{kk})}{\det(\mathbf{B})} + \sum_{i=1, i \neq k}^{m} I_{yi}(l) b_i + K_y \quad (22)$$

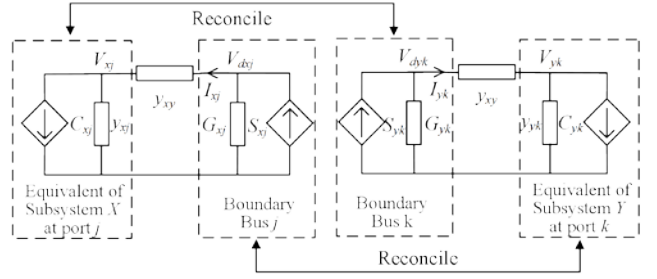

Fig. 5. The equivalent representation of the system as shown in Fig. 3 (b)

where $\mathbf{B} = \begin{bmatrix} y_{y1} & h_{y21} & \cdots & h_{ym1} \\ h_{y12} & \ddots & & h_{ym2} \\ \vdots & & & \vdots \\ h_{y1m} & h_{y2m} & \cdots & y_{ym} \end{bmatrix}$, $b_i = \dfrac{(-1)^{i+k} \det(\mathbf{M}_{ik})}{\det(\mathbf{B})}$,

and $K_y$ denotes the constant factor resulting from active sources in subsystem $Y$.

Expanding on the $j$th and $k$th columns, respectively, allows the cofactors to depend on self and coupling admittances only. As is shown next, the exclusion of $I_x(l+1)$ and $I_y(l)$ terms from the cofactor terms will allow the cofactor terms to form the row entries in the **W** matrix.

By substituting Eq. (21) and Eq. (22) into Eq. (19), the left-hand side of the equation becomes:

$$\left( \frac{1}{y_{xy}} + \frac{1}{G_{xj}} + \frac{\det(\mathbf{M}_{jj})}{\det(\mathbf{A})} \right) I_{xj}(l+1) + \sum_{i=1, i \neq j}^{n} I_{xi}(l+1) a_i \quad (23\text{-a})$$

and the right-hand side of the equation becomes:

$$\left( \frac{\det(\mathbf{M}_{kk})}{\det(\mathbf{B})} - \frac{1}{G_{xj}} \right) I_{yk}(l) + \sum_{i=1, i \neq k}^{m} I_{yi}(l) b_i + K \quad (23\text{-b})$$

where $K = K_y - K_x$.

Therefore, it is observed that a linearly weighted algebraic sum of all port currents of subsystem $X$ updated to $l+1$ is equal to an algebraic sum formed by a single *difference term*

$$\Delta_k = \left( \frac{\det(\mathbf{M}_{kk})}{\det(\mathbf{B})} - \frac{1}{G_{xj}} \right) \quad (24)$$

multiplying the port $k$ boundary bus current of subsystem $Y$ at iteration $l$ and a set of *direct terms* $b_i$ ($i \neq k$) that link to the other $m$-1 boundary bus currents of subsystem $Y$ at iteration $l$.

The process of expanding Eq. (19) for all ports of subsystem $X$ results in $n$ independent equations like Eq. (23). Cramer's rule could again be applied to solve for each $I_{xj}(l+1)$, $j \in [1, \ldots n]$; and doing so forms the $n$ row entries of the **W** matrix due to subsystem $X$. This process can be repeated for each subsystem until all row entries in **W** are filled in.

It is not the purpose of this paper to pursue forming the **W** matrix as a solution method, even though the process described in this subsection would enable such an approach provided that the constant vector **G** of Eq. (5) were derived as well. Instead, this subsection will conclude by reciting three basic remarks about what to expect from constructing the **W** matrix for the PUR solution and then the following subsection will set the degree of freedom (i.e. $G_{xj}$) of Eq. (23) to explicitly define the PGNME procedure and to permit an evaluation of the procedure's impact on the spectral radius.

*Rule 1*: Each row of **W** will contain zeros in every position multiplying a port current of the subsystem represented by the row. This will form a set of square block matrices filled with zero entries along the main diagonal of **W**. Zeros along the diagonal of **W** are a fundamental property of the PUR method.

*Rule 2*: Each row of **W** will contain a *difference term* similar to Eq. (24) in every position multiplying a port current that updates the subsystem represented by the row. The "updating" port currents belong to the ensemble of subsystems directly connected to one or more of the $n$ boundary buses of the subsystem represented by the row. It will be shown in section 3.3.2 that these difference terms are where PGNME is defined and the benefits accrued, subject to limitations imposed by the "residue" which will also be defined in section 3.3.2.

*Rule 3*: Each row of **W** will contain a *direct term* similar to $b_i$ of Eq. (22) for every bus current that does not update (the updating currents are covered by Rule 2) belonging to the ensemble of subsystems directly connected to one or more of the $n$ boundary buses of the subsystem represented by the row. This is because every node voltage of a subsystem connected to the subsystem represented by the row directly influences the bus current being iterated when Eq. (9) and Eq. (10) are solved at each iteration step. It will be shown that the direct terms can impose additional limitations on the benefits of PGNME.

### 3.3.2 Modifying the difference term to define PGNME

Combining the aforementioned rules on the general structure of the iteration matrix for PUR solution, **W**, and its convergence properties discussed in Eq. (5) to Eq. (7) in Section 2.4, in the following subsection, we shall focus on demonstrating the fundamental concept and derivation of PGNME preconditioning and how it can effectively optimize the convergence properties (i.e. reduce the spectral radius $\rho(\mathbf{W})$).

According to Gershgorin's Theorem, Theorem 2.1 in [23] states that:

Let $A$ be a complex $n \times n$ matrix with entries $a_{ij}$, every eigenvalues of matrix $A$ satisfies:

$$\left|\lambda - a_{ii}\right| \leq \sum_{i \neq j}\left|a_{ij}\right|, i \in \{1,2,...,n\} \quad (25)$$

**Lemma 1**. For **W**, the following derivation is valid:

$$\left|\lambda\right| \leq \sum_{i \neq j}\left|a_{ij}\right|, i \in \{1,2,...,n\} \quad (26)$$

*Proof.* From Rule 1, all entries along the diagonal of iteration matrix **W** are zero, therefore all $a_{ii}$ are zero. Thus based on Eq. (25), Lemma 1 can be derived.

Lemma 1 indicates that the upper bound of the spectral radius of **W** can be determined by the sum of the absolute values of the non-diagonal entries on the $i$-th row. This leads to the result that if the absolute values of the off-diagonal entries can be reduced, the upper bound of the spectral radius can also be reduced.

Based on Rule 2 and Rule 3, the off-diagonal entries of **W** contain broadly two types of entries, difference terms (described by Rule 2) and direct terms (described by Rule 3). In addition, each difference term is in a form similar to $\Delta_k$ of Eq. (24), while each direct term is in the form similar to $b_i$ of Eq. (22). This suggests that although one may not be able to find the precise minimum and maximum eigenvalues for iteration matrix, the estimated boundaries of the spectrum may still be obtained by exploring the general structures of these two terms. Per Lemma 1, the upper limit of $\rho(\mathbf{W})$ can be assessed and tuned to improve the convergence properties of the PUR solution.

Although the port dependencies within a multi-port subsystem can be dependent on various factors and system specifications, an estimate that generally holds true is that when the resistance distances (as defined in [24], [25]) between ports become greater, the port dependencies become weaker. In other words, when the size of the system is sufficiently large compared to the number of the ports within the system, then it becomes naturally likely that the port coupling coefficients are small. If this assumption is valid, the off-diagonal entries of the multi-port equivalent matrices described in Eq. (14) become insignificant; therefore, it can be assumed the multi-port equivalent matrices are row diagonally dominant in the form of:

$$\left|\mathbf{m}_{ii}\right| \geq \sum_{j \neq i}\left|\mathbf{m}_{ij}\right| \quad (27)$$

where **m** denotes the multi-port equivalent matrices for a subsystem.

**Lemma 2**. If all the multi-port equivalent matrices within a partitioned system are row diagonally dominant, then by performing the so-called PGNME preconditioning, the difference terms within **W** matrix can be reduced to residues.

*Proof.* According to Corollary 2.3 and Theorem 2.6 in [34], the determinant of a row diagonally dominate matrix can then be approximated by the products of its diagonal elements. In other words, for the difference term $\Delta_k$ of Eq. (24), we have

$$\det(\mathbf{B}) = \prod_{i=1}^{m} y_{yi} + \tau \quad (28\text{-}a)$$

and

$$\det(\mathbf{M}_{kk}) = \frac{1}{y_{yk}} \prod_{i=1}^{m} y_{yi} + \delta \quad (28\text{-}b)$$

where $\tau$ and $\delta$ are small "errors" incurred in estimating the determinants using the product of the entries on the main diagonal of each matrix, and the upper bounds of these errors can be calculated using Theorem 2.6 as well. Based on Eq. (28) and Eq. (24), the difference term can now be shown to be

$$\Delta_k = \left(\frac{\det(\mathbf{M}_{kk})}{\det(\mathbf{B})} - \frac{1}{G_{xj}}\right) = \left(\frac{1}{y_{yk}} - \frac{1}{G_{xj}}\right) + \varepsilon \quad (29)$$

where the residue term becomes an error $\varepsilon$ due to the diagonal approximation. The question is, how significant is the magnitude of $\varepsilon$? If $\varepsilon$ is normalized by $y_{yk}^{-1}$, it can be shown that:

$$\frac{\varepsilon}{y_{yk}^{-1}} = \frac{y_{yk}\delta - \tau}{\prod y_{yi} + \tau} = \frac{\delta}{\frac{1}{y_{yk}}\left(\prod y_{yi} + \tau\right)} - \frac{\tau}{\prod y_{yi} + \tau} \quad (30)$$

It is not hard to realize from Eq. (30) that if the error terms $\tau$ and $\delta$ are small compared to the products they are summed with in Eq. (28-a) and Eq. (28-b), respectively, then the residue in Eq. (29) will also be small compared to the other admittances. Eq. (30) is useful for testing the assertion of row diagonal dominance during the analysis of the numerical simulation results.

Eq. (29) indicates that the difference term can be approximated by $\left(y_{yk}^{-1} - G_{xj}^{-1}\right)$ when all the multi-port equivalent matrices resulted from the partitioning are row diagonally dominant, and by setting $G_{xj}^{-1} = y_{yk}^{-1}$ or equivalently $G_{xj} = y_{yk}$, the difference term $\Delta_k$ can be reduced to a mere residue which suggests that the upper bound of $\rho(\mathbf{W})$ is also reduced according to Lemma 1. This allows PGNME to be formally defined as a preconditioning step where the boundary bus admittance is set to $G_{xj} = y_{yk}$, where $y_{yk}$ is the self-admittance of the $k$th port of subsystem $Y$ that shares a cut boundary bus with the $j$-th port of subsystem $X$ as shown in Fig. 4, so that the first binomial term of Eq. (29), namely, the *PGNME difference term*, vanishes and the difference term is left with a *residue term*. (In Fig. 4 PGNME would require $G_{yk} = y_{xj}$ also.) This decision to settle the degree of freedom this way is prompted by previous literature [35-36]. It will be shown by example in section 4 that PGNME will reduce the spectral radius; and in some special cases reduce it to zero.

But Eq. (29) also shows that in general the application of PGNME alone should not be expected to reduce the spectral radius or its upper bound to zero, because the second binomial term of Eq. (29) ε, the "residue" of PGNME, is not automatically zero. Instead, an example in section 4 will show that the off-diagonal terms of the $\mathbf{W}$ matrix formed by the residues and the direct terms are proportional to port coupling admittances (i.e., the $h_{ij}$ of the $\mathbf{A}$ and $\mathbf{B}$ matrices). Therefore, the effectiveness of PGNME applied alone will depend upon the critical condition that the size of the system is sufficiently large compared to the number of the ports within the system, so that the port dependencies within each partition can be considered weak. We shall demonstrate this observation in the results shown in section 6.

Note that during the simulation it is assumed that system topologies and configurations are approximately stationary, therefore the partitioning schemes, equivalent admittances and the modified nodal admittance matrices stay approximately unchanged during the simulation procedure. However, it is not necessary that they stay identically stationary because as long as the spectral radius is less than one, then PUR will iterate until a solution converges. In fact, if the spectral radius starts small due to initializing with PGNME, then small changes to the system may not cause an increase in the number of iterations needed to converge until the changes become significant. If the structure of the system has drastically changed due to large-scale events, then the simulation may need to be paused and reinitialized/repartitioned in order to maintain the performance of the simulation, which is easily recognized by tracking the iteration count.

## 3.4 Complexity analysis of PGNME

Considering a generic power system network in the form of Fig. 2 decomposed into $p$ subgraphs through a total number of $m$ branch cuts, under the assumption that the system is decomposed to subsystems that are roughly about the same size, and each partition roughly contains a total number of $n$ buses plus $2m/p$ boundary buses resulted from the partitioning process, then the size of the original network can be represented as $N=p\times n$. Define $r$ as the number of iterations until the global convergence, $t_c$ as the communication cost to send/received MPI messages, and $q$ as the total number of processing units (in our context, cores) that the calculation is being distributed to ($q>1$), the total computation cost of PGNME can be mathematically quantified as follows:

### 3.4.1 For factorization

Taking a conventional sparse matrix solution LU decomposition for example, the computational complexity of factorizing an $N\times N$ nodal admittance matrix used in Eq. (3) is $O(N^3)$ or $O(p^3 n^3)$. Using PGNME, all the factorization become completely independent. Since each partition now consists of a $(n+2m/p)\times(n+2m/p)$ nodal admittance matrix with boundary buses added, the total cost of factorizing all $p$ partitions using a total number of $q$ cores can be represented as:

$$\phi_F = k_1(n+2m/p)^3 [p/q] \quad (31)$$

where $k_1$ is the scaling coefficient. For a large system, it can be assumed that the total number of boundary buses $2m$ are far less than the total number of buses $n$. Under this assumption, Eq. (31) demonstrates that for factorization, PGNME scales as $O(p/q)$. While this complexity reduction seems significant, this factorization is only a small portion of the total simulation. The benefit from this portion will be more realizable when the structure of the nodal admittance matrix changes often during simulations such as load shedding, switching events, etc.

### 3.4.2 For solving

For LU methods, the complexity for solving the original problem is $O(N^2)$, while using PGNME, the solving is independent but there is also an iterative updating strategy. Therefore, for $p$ number of partitions, the total computational cost of complexity for each iteration can be derived as $\phi_L = k_2(n+2m/p)^2[p/q]$ where $k_2$ is the scaling coefficient. Since the above process needs to be redone for a total of $r$ times, and consider the overhead of data communications among the processors, the total cost of the solving part of PGNME can be represented as:

$$\phi_L = k_2(n+2m/p)^2[p/q]r + k_3 t_c(m/p)r \quad (32)$$

where $k_3$ is the scaling coefficient for the communication overhead. Under the assumption that $m \ll n$ and the communication cost $t_c$ can be considered trivial, Eq. (32) demonstrates that for the solving stage of PGNME, the computational cost scales as $O([p/q]r)$.

Based on Eq. (31) and Eq. (32), it can be observed that the main computational advantage of PGNME as a Jacobian-like decomposition algorithm is that all the subgraphs can be solved independently. This translates to an almost linear scaling of the computational cost of PGNME with the number of subsystems while the original network solution computation suffers from a superlinear scaling.

However, it should also be noted in Eq. (32) that the computational cost also scales linearly with the number of iterations. Therefore, controlling the convergence of the overall reconciliation process also plays a critical role in the determination of the computational performance of PGNME.

## 4 EXAMPLES WITH PGNME

In this section, we focus on demonstrating how the general concept of PGNME developed in Section 3 can be utilized in parallelizing practical power system simulation problems. We start with a very simple two-subgraph problem that has been discussed in literature and show that the general derivation of PGNME can be applied and leads to the same outcome that was concluded in the literature. Then we will present a more general multi-subgraph, multi-port example as the proof-of-concept to show the convergence improvement effect of PGNME.

### 4.1 A two-subgraph, single-port example

In this example, we consider the decomposition for a simple 7-node system shown in Fig. 6 (a). After the branch cut at $y_{ab}$, two subsystems can be formed, namely, subsystem 1 and subsystem 2. Boundary bus 1 and 2 are created and attached to the terminal of subsystem 1 and 2 respectively. Unlike the multi-port case considered in the general PGNME derivation, in this example both subsystem 1 and 2 contain only one port. Therefore, this example can be considered a simplified application of PGNME without multi-port equivalent calculation, or just simply PGN.

In order to develop the analytical model of the reconciliation process for this two-subgraph system, the equivalent representations of subsystem 1 and 2 need to be identified in the first place. Since there are no other ports existing in both of the subsystems, the equivalent representation of the subsystems can be obtained by following the steps defined in Section 3.3.1. Once the equivalent current sources and admittances are determined, we can construct the complete equivalent representation for each subproblem as shown in Fig. 6 (b). Based on Eq. (15), the equivalent matrix for subsystem 1 and 2 can be formulated as

$$I_a = \mathbf{A} V_a + C_a \quad (33\text{-}a)$$

and

$$I_b = \mathbf{B} V_b + C_b \quad (33\text{-}b)$$

respectively, where $\mathbf{A} = y_a$, $\mathbf{B} = y_b$.

Compare Eq. (33) with the general representation considered in Eq. (15), it can be observed that since each subproblem only contains one single-port, the linear port dependency $h_{xji}$ ($i, j \in [1,...n], i \neq j$) doesn't exist in this example, therefore, the equivalent current sources (denoted by $C_a$ and $C_b$) are solely determined by the internal current sources in subsystem 1 and 2 respectively.

Based on Eq. (23), the iterative process for this example can be written:

$$\left(\frac{1}{y_{ab}} + \frac{1}{G_1} + \frac{1}{y_a}\right) I_a(l+1) = \left(\frac{1}{y_b} - \frac{1}{G_1}\right) I_b(l) + k_a \quad (34\text{-}a)$$

for subsystem 1. And

$$\left(\frac{1}{y_{ab}} + \frac{1}{G_2} + \frac{1}{y_b}\right) I_b(l+1) = \left(\frac{1}{y_a} - \frac{1}{G_2}\right) I_a(l) + k_b \quad (34\text{-}b)$$

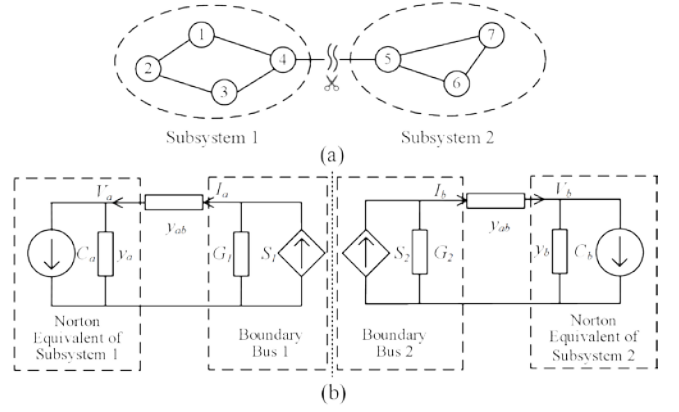

Fig. 6. A two-subgraph, single-port example is shown in (a) while its equivalent representation is shown in (b)

for subsystem 2, where $k_a$ and $k_b$ denote the constant factors. It is noted that the residues of the difference terms in Eq. (34) are zero because the $\det(M_{kk}) = 1$ and $\det(B) = y_b$ in Eq. (34-a) or $y_a$ in Eq. (34-b). This occurs whenever the subsystem on the opposite side of the cut edge is a single-port subsystem because the Norton equivalent looking into this port is an exact reduced order model of the subsystem.

Rewriting Eq. (34) in the first-order linear stationary form defined in Eq. (5) results in Eq. (35) below:

$$\begin{bmatrix} I_a(l+1) \\ I_b(l+1) \end{bmatrix} = \begin{bmatrix} 0 & \lambda_1 \\ \lambda_2 & 0 \end{bmatrix} \cdot \begin{bmatrix} I_a(l) \\ I_b(l) \end{bmatrix} + \begin{bmatrix} k_a \\ k_b \end{bmatrix} \quad (35)$$

where $\lambda_1 = \left(y_b^{-1} - G_1^{-1}\right)\left(y_{ab}^{-1} + G_1^{-1} + y_a^{-1}\right)^{-1}$ and

$\lambda_2 = \left(y_a^{-1} - G_2^{-1}\right)\left(y_{ab}^{-1} + G_2^{-1} + y_b^{-1}\right)^{-1}$.

It is then obvious that the spectral radius of the iteration matrix can be obtained by inspection of Eq. (35) to be

$$\rho(\mathbf{W}) = \sqrt{\lambda_1 \lambda_2} \quad (36)$$

It can be clearly observed from Eq. (35) that in this example, the convergence property of the iterative process is directly determined by the values of which can be reduced to 0 by setting the boundary admittances $G_1$ and $G_2$ contained in the boundary buses to the equivalent admittance $y_a$ and $y_b$. This outcome justifies our conclusion made on the PGNME difference term in Section 3.3.2. Meanwhile, this example and the general usage of PGN should be considered only a "special case" as there are no port dependencies existing within the subsystems. The direct effect of this simplification is that the general multi-port equivalent matrix in Eq. (14) is reduced to a scalar, and the direct terms $a_i$ and $b_i$ in Eq. (23) also become zero. Therefore, we could easily manipulate the iteration matrix $\mathbf{W}$ and set its spectral radius to zero for the ideal case of instantaneous convergence.

While this example is sufficiently intuitive to explain the basic motivation and procedure for PGNME preconditioning, PGN is of little practical use by itself because it is strictly applicable to a graph partitioned into only two subgraphs. Even the average personal computer has more than two cores available for the opportunity to parallelize the solution, but truly large problems that could be run on a general purpose high-performance cluster computer can benefit from hundreds or thousands of partitions, as will be demonstrated by the simulation results reported in section 6. Therefore, the previous literature [35-36] describing

PGN is inapplicable and unfeasible for practical problems of any size when the system is decomposed into $p$ subsystems, where $p > 2$. In the next subsection, we will consider a more complicated example which represents the practical challenge in decomposing large-scale power system networks.

### 4.2 A multi-subgraph, multi-port example

In this example, we consider a system as shown in Fig. 7 where three two-port subsystems are tied together by three branches that will become cut edges in the decomposed system. Unlike the previous example where no port dependencies exist, in this example, each subproblem contains two ports that are directly coupled; thus PGN becomes immediately inapplicable. In the following discussion, we will focus on demonstrating how to apply the general derivation shown in Eq. (23) to evaluate the convergence propery of the PUR solutions for this example and further examining the three basic rules summazied in subsection 3.3.1 and the effect of PGNME preconditioning on the spectrula radius upper bound reduction illustrated in subsection 3.3.2.

To start with, the equivalent representations of each subsystem can be calculated using the steps defined in subsection 3.3.1. Using the notations defined in Fig. 7, it can be assumed that the multi-port equivalent matrices can be derived as $\mathbf{A}_x = \begin{bmatrix} y_{x1} & h_{x21} \\ h_{x12} & y_{x2} \end{bmatrix}$, $\mathbf{A}_y = \begin{bmatrix} y_{y1} & h_{y21} \\ h_{y12} & y_{y2} \end{bmatrix}$ and $\mathbf{A}_z = \begin{bmatrix} y_{z1} & h_{z21} \\ h_{z12} & y_{z2} \end{bmatrix}$ for subsystem X, Y, and Z respectively. Writing six independent equations based on Eq. (23) and then solving for the six bus currents at iteration l+1, the iterative process can be represented by Eq. (37) below:

$$\begin{bmatrix} I_{x1}(l+1) \\ I_{x2}(l+1) \\ I_{y1}(l+1) \\ I_{y2}(l+1) \\ I_{z1}(l+1) \\ I_{z2}(l+1) \end{bmatrix} = \begin{bmatrix} 0 & 0 & \alpha_{y11} & a_{y12} & a_{z11} & \alpha_{z12} \\ 0 & 0 & \alpha_{y21} & a_{y22} & a_{z21} & \alpha_{z22} \\ b_{x11} & \beta_{x12} & 0 & 0 & \beta_{z11} & b_{z12} \\ b_{x21} & \beta_{x22} & 0 & 0 & \beta_{z21} & b_{z22} \\ \gamma_{x11} & c_{x12} & c_{y11} & \gamma_{y12} & 0 & 0 \\ \gamma_{x21} & c_{x22} & c_{y21} & \gamma_{y22} & 0 & 0 \end{bmatrix} \begin{bmatrix} I_{x1}(l) \\ I_{x2}(l) \\ I_{y1}(l) \\ I_{y2}(l) \\ I_{z1}(l) \\ I_{z2}(l) \end{bmatrix} + \mathbf{K}$$

(37)

By inspection of Eq. (37) it is obvious that the iteration matrix $\mathbf{W}$ conforms to Rule 1 reported in subsection 3.3.1 which states that the block matrices on the main diagonal of $\mathbf{W}$ are zero matrices.

For the sake of brevity, in the following discussion, we mainly focus on studying the iterative process involving subproblem X as an example of the overall iterative process described by Eq. (37). However, the iterative processes for the other subproblems can be formulated and evaluated in the same manner.

The iterative process of subproblem X is governed by the first two rows of the iteration matrix in Eq. (37). Based on the detailed formulations for each entry (as shown in the appendix), we can categorize the entries included in the off-diagonal block matrices of $\mathbf{W}$ into two types: entries that contain difference terms (labelled by Greek letter $a$) and entries that contain only direct terms (labelled by English letter $a$). The difference terms are associated with port currents $I_{y1}$ and $I_{z2}$ which are directly connected to subsystem X as shown in Fig. 7. The direct terms are associated with port currents $I_{y2}$ and $I_{z1}$ which are the other two port currents associated with the two subsystems directly connected to subsystem X that can affect the boundary conditions of subsystem X through the port dependencies existing in subsystem Y and Z respectively. This result precisely conforms to Rule 2 and Rule 3 stated in subsection 3.3.1. Another observation we can make based on the detailed formulations of each entry found in the appendix is that both the direct terms and the residues of the difference terms are proportional to products of the port coupling coefficeints (i.e., $h$ coefficients in Eq. (14)). Therefore, we can conclude that the convergence properties of the PUR solution when applied to a multi-subgraph, multi-port system are directly affected by the port dependencies within each subproblem and the ensemble of subproblems it is connected to.

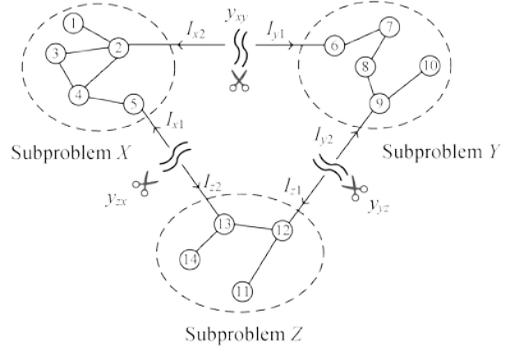

Fig. 7. Network decomposition and boundary bus creation of a 3-parititoned 14-node system

Applying PGNME preconditioning to the iteration process of this example, based on the procedures illustrated in subsection 3.3.1, the boundary bus admittances contained in each boundary bus $G_{x1}, G_{x2}, …, G_{z2}$ can be set accordingly to $y_{z2}, y_{y1}, …, y_{x1}$. Following this procedure, the iteration matrix shown in Eq. (37) can be reduced to:

$$\mathbf{W'} = \begin{bmatrix} 0 & 0 & \varepsilon_{y11} & a_{y12} & a_{z11} & \varepsilon_{z12} \\ 0 & 0 & \varepsilon_{y21} & a_{y22} & a_{z21} & \varepsilon_{z22} \\ b_{x11} & \varepsilon_{x12} & 0 & 0 & \varepsilon_{z11} & b_{z12} \\ b_{x21} & \varepsilon_{x22} & 0 & 0 & \varepsilon_{z21} & b_{z22} \\ \varepsilon_{x11} & c_{x12} & c_{y11} & \varepsilon_{y12} & 0 & 0 \\ \varepsilon_{x21} & c_{x22} & c_{y21} & \varepsilon_{y22} & 0 & 0 \end{bmatrix}$$ (38)

It can be clearly observed that PGNME preconditioning effectively reduces the upper bound of $\rho(\mathbf{W'})$ compared with $\rho(\mathbf{W})$ under the assumption that $\mathbf{A}_x$, $\mathbf{A}_y$ and $\mathbf{A}_z$ are all row diagonally dominant.

## 5 PGNME IMPLEMENTATION

In this section, we brief explain how to map the structure of the proposed PGNME method onto the parallel computing architecture and develop a simulation program to implement, test and evaluate the performance of PGNME preconditioning when applied to parallel power system dynamic simulation.

### 5.1 Hardware and software setup

The HPC platform used in this study is a state-of-art

HPC cluster "Shadow II" available to us at the High Performance Computing Collaboratory at Mississippi State University. Shadow II consists of 110 nodes, each node containing 512 GB of RAM and 2 Intel E5-2680 v2 Ivy Bridge processors, which are each 10 core and operate at 2.8 GHz. The communication system is FDR InfiniBand (56 Gb/s).

The implementation of the reconciliation processes among subproblems is through the Intel® MPI Library based on the MPI 3.0 standard [37-38]. This MPI/OpenMP approach uses an MPI model for communicating between nodes while utilizing groups of threads running on each computing node in order to take advantage of multi-core/many-core architectures. A more detailed documentation of MPI 3.0 standard can be found in [38].

To map the subproblems into the hardware, in the results section the number of MPI processes required is always equal to the number of partitions(subproblems) plus one (an additional process needs to be created to manage the overall reconciliation process). Within all simulations there is a simple one to one mapping of processes to a physical core. While this is the current setup to examine the performance of PGNME, alternative reconciliation structures such as point-to-point and hierarchical reconciliation strategies could be considered in future work.

### 5.2 Graph partitioning

Based on the aforementioned discussion, a power system network presented in the form of a graph needs to be first decomposed into subgraphs for parallel simulation during the initialization stage. The graph-based partitioning tool adopted in this study is the well-known program hMeTiS [31]. More specifically, the unweighted, multi-level based hMeTiS partitioning algorithm is used for all the case studies as the main purpose for the graph partitioner is to generate roughly even size subgraphs and minimize the inter-connections between subgraphs for the purpose of load balancing and inter-node communication reduction. While our aforementioned discussion indicate that an appropriate partitioning algorithm could potentially help extend the breakpoint of PGNME to gain more parallelism, it is not the focus of this paper and thus is not considered.

### 5.3 Subgraph solver

After the decomposition, each subproblem still needs a numerical solver to solve the partitioned subgraph. A commonly used sparse matrix solver SuperLU_sequential [40-41] is deployed in this paper. Note that SuperLU_sequential is chosen for its simplicity and robustness for implementation, but any existing sequential linear solver could be utilized with the proposed parallel updating strategy.

## 6 SIMULATION RESULTS

Simulation results are provided in this section to assess the performance of PGNME preconditioning in terms of accuracy, computational efficiency, and scalability. More specifically, two case studies are presented to demonstrate the operation and performance of PGNME when tackling the main challenge of this paper: parallel dynamic simulation for large-scale systems. We will start with case study 1 where a power network of roughly 28,000 buses which is on the same scale of the Eastern Interconnection (EI) is simulated. In case study 2, an arbitrary system that consists of over 258,000 buses is studied. We argue that even a power grid of such scale does not exist in any practical application yet, by examining the performance of PGNME on the extreme scale, we can directly observe the scalability of the proposed approach and explore both the algorithmic and hardware limitations.

### 6.1 Case study 1

The performance of PGNME is firstly tested on an arbitrary network that contains roughly 28,647 buses which approximates the size of the current EI system. In order to emulate the ever-changing component dynamics, arbitrary excitations are also added to the generator and load buses. By controlling the magnitude and the frequency of the excitations, it becomes possible for us to observe the performance of PGNME when the system is within steady-state operating condition and when the system is subject to wide-area, quick spreading perturbations such as faults or other dynamic events.

To start with, we compared the performance of un-preconditioned PUR and PGNME with regards to the average number of iterations (*a.n.i*) when applied to the system under study. The averaging process can be described as:

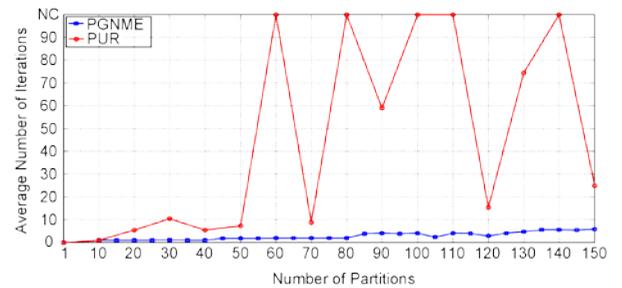

Fig. 8. The performance comparison between un-preconditioned PUR and PGNME under slow-dynamics in terms of averaged relaxation iterations

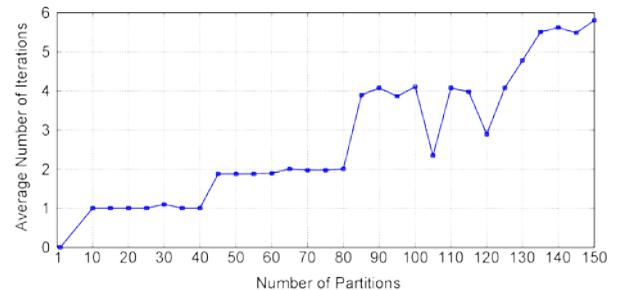

Fig. 9. The average number of iterations using PGNME for test system 1

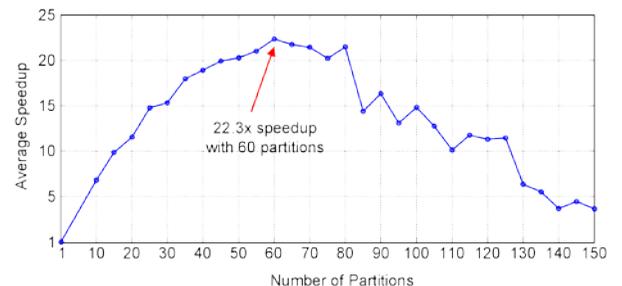

Fig. 10. The overall speed up obtained by utilizing PGNME for test system 1

$$a.n.i = \frac{N_{iteration}}{N_{step}} \quad (39)$$

where $N_{iteration}$ refers to the total number of iterations counted from the beginning of the simulation to the end of the simulation and $N_{step}$ denotes the total number of time steps. Note that the one iteration count means the MPI communication has happened and the convergence has been checked2. Through the averaging process, we are able to make an overall more accurate and more comprehensive observation of the performance of PGNME when the system is partitioned in different ways. The comparison is shown in Fig. 8. It can be clearly observed that while PUR suffers from an intractable number of iterations and in some cases, do not even converge within 100 iterations (those cases are marked as "NC" in Fig. 8), PGNME can always maintain improved convergence rates under different partitioning schemes. Specifically, Fig. 9 has shown that before the number of partitions reach 40, PGNME is able to achieve convergence in one iteration which can be considered "instantaneous". When the number of partitions increases, it can be observed that after 40 partitions, roughly one extra iteration is added in the convergence process. This effect becomes more evident after 80 partitions, where the *a.n.i* for PGNME quickly exceeds two and stays over four until the end of the testing; however, compared with its control, the un-preconditioned PUR, PGNME always provides a considerable reduction in the *a.n.i* throughout the comparison.

This observation directly validates the previous prediction made based on the general structure of **W** and Lemma 2, which suggests that once the total number of partitions becomes sufficiently large, the assumption underpinning PGNME becomes less and less dominant as the coupling parameters becomes generally more and more significant compared with the equivalent admittance at the port. In other words, the multi-port equivalent matrices for the partitioned subsystems cannot be guaranteed to be adequately independent of loading imposed by the connection to other subsystems at the other ports on the boundary of the same subsystem. This explains the extra iterations gained at 40 and 80 partitions as the iterative process needs to utilize additional iterations to compensate the effect of close subproblem internal couplings. Therefore, it can be concluded that by utilizing PGNME, an iterative process with greatly improved and more traceable convergence properties can be obtained to decompose and solve the nodal network equations described by Eq. (3) in parallel. Although the convergence enhancement effect of PGNME can no longer be guaranteed when the sizes of the subsystems become significantly large, it still offers a great advantage and the necessary computational capability especially when the size of the original system is sufficiently large, which is the challenge we aimed to address, and the total number of processors (subsystems) are limited, which is always the case for actual applications. We will demonstrate this conclusion particularly in the next case study.

The computational speed-up gained by adopting PGNME is shown in Fig. 10. It can be observed that with PGNME preconditioning, a roughly 22 times speed up can be obtained with 60 processes.

## 6.2 Case study 2

In case study 2, we study a much larger network to investigate and ensure PGNME's performance at larger scales, which is our primary concern and motivation for this paper. An arbitrary 258066 bus system was created for this case study. We took a similar approach as performed in literature such as [15], which assembles the arbitrary power system by spawning a group of randomly variated standard IEEE-118 bus systems, and then creating random inter-connections between these variated parts to form the properly scaled system that can be used for the test. The size of the generated system matches the potential size of future EU mega-grid envisioned in [42]. Similar to the first case study, the results presented in this case study are all obtained under arbitrary dynamic profiles.

The average number of iterations PGNME takes to converge for test system 2 with different number of partitions are shown in Fig. 11. The simulation results indicate that when the size of the system grows (roughly 10 times larger than test system 1), the performance of PGNME extends accordingly. More specifically, PGNME is capable of maintaining an average number of iterations of 2 until 140 partitions. The overall average speed up for case study 2 is shown in Fig. 12. The peak of the speedup curve is at 160 partitions with a maximum 101 times speed up. Comparing with the subsequent partitions, it can be clearly observed that the speed up declines after 160 partitions due to the limitations of the hardware platform, particularly the communication overheads, since 180-240 partitions maintain roughly the same convergence properties (with an average of 3 iterations). This further proves that the SRR technique included in PGNME is able to effectively improve the convergence properties of the iterative process especially when the size of the system is adequately large compared with the number of partitions. It also suggests that PGNME is capable of fully utilizing the computational resources used in this case study to gain speedup, because if there were no such hardware limitations, the speedup curve shown in Fig. 12 would have kept rising until 240 partitions at which point the iteration count begins to rise unacceptably as shown in Fig. 11. This clearly indicates that PGNME, as a domain decomposition algorithm, is a natural fit for applications that require dynamic simulations of massive scale power systems.

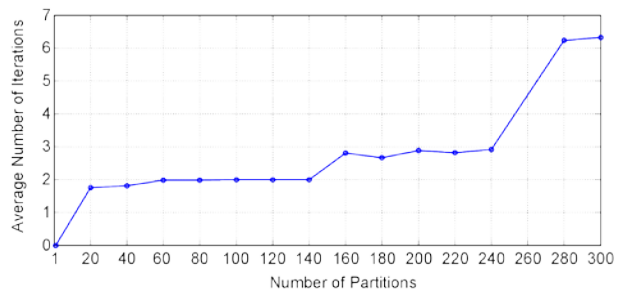

Fig. 11. The average number of iterations PGNME takes to converge for test system 2

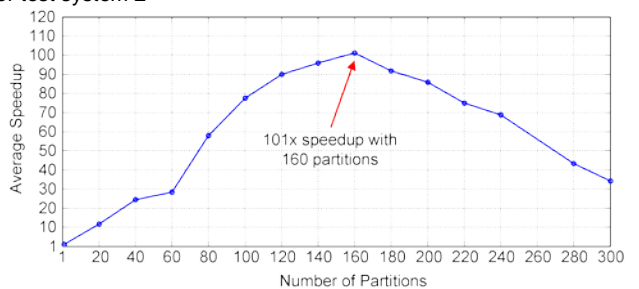

Fig. 12. The overall speed up for test system 2 using PGNME

# 7 CONCLUSION

This paper proposes a novel network decomposition technique, namely, PGNME and demonstrates its advantages when applied to large-scale power system simulations. By greatly improving the convergence property of the iterative process, PGNME can be seen as the algorithmic enabler for solving a variety of emerging problems in the study of large-scale, interconnected power systems, such as look-ahead dynamic simulation, real-time operational enhancement, dynamic security assessment, and predictive protection and control using high-performance computing.

While this paper focuses on describing and validating a novel method for decomposing and solving Eq. (3), future work will further address the complete decomposition and self-consistent solution of all the equations included in both Eq. (2) and Eq. (3). The performance limitation of PGNME will also be investigated. As demonstrated in the theoretical derivation and simulation results, PGNME becomes less likely to maintain the convergence properties when the sizes of the subsystems become relatively small. The consequence of this limitation, which is shown in Section 6, is that the average number of iterations start to rise and can no longer stay within the expected range when the number of partitions exceeds a certain "threshold". Therefore, it becomes imperative to postpone or even overcome this performance threshold in order to gain massive parallelism. Another aspect of the performance validation will include a comprehensive comparison between the proposed DD approach and the common fine-grained data parallel solver such as PETSc to highlight the difference between a general-purpose parallel sparse linear solver package and a domain-specific preconditioning/reconciliation technique. These issues and challenges will be investigated further in future works.

## REFERENCES

[1] Kundur, P., N.J. Balu, and M.G. Lauby, Power system stability and control. Vol. 7. 1994: McGraw-hill New York.
[2] Stott, B., Power system dynamic response calculations. Proceedings of the IEEE, 1979. 67(2): p. 219-241.
[3] U.S Department Of Energy, The Future of the Grid: Evolving to Meet America's Needs. 2014.
[4] Office of Electric Transmission and Distribution, Grid 2030: A national Vision for electricity's second 100 years. 2003, U.S Department of Energy.
[5] Eto, J.H. and R.J. Thomas, Computational needs for the next generation electric grid. U.S Department of Energy, 2011.
[6] Huang, Z., S. Jin, and R. Diao. Predictive dynamic simulation for large-scale power systems through high-performance computing. in High Performance Computing, Networking, Storage and Analysis (SCC), 2012 SC Companion:. 2012. IEEE.
[7] Huang, Z., et al. Capturing real-time power system dynamics: Opportunities and challenges. in 2015 IEEE Power & Energy Society General Meeting. 2015.
[8] Toselli, A. and O.B. Widlund, Domain decomposition methods: algorithms and theory. Vol. 34. 2005: Springer.
[9] Aristidou, P., D. Fabozzi, and T.V. Cutsem, Dynamic Simulation of Large-Scale Power Systems Using a Parallel Schur-Complement-Based Decomposition Method. IEEE Transactions on Parallel and Distributed Systems, 2014. 25(10): p. 2561-2570.
[10] Saad, Y., Iterative methods for sparse linear systems. 2003: Siam.
[11] Jin, S., et al. Parallel implementation of power system dynamic simulation. in 2013 IEEE Power & Energy Society General Meeting. 2013. IEEE.
[12] Diao, R., et al. "On Parallelizing Single Dynamic Simulation Using HPC Techniques and APIs of Commercial Software." IEEE Transactions on Power Systems 32.3 (2017): 2225-2233.
[13] Jin, S., et al. "Comparative implementation of high performance computing for power system dynamic simulations." IEEE Transactions on Smart Grid 8.3 (2017): 1387-1395.
[14] Jalili-Marandi, V. and V. Dinavahi, SIMD-Based Large-Scale Transient Stability Simulation on the Graphics Processing Unit. IEEE Transactions on Power Systems, 2010. 25(3): p. 1589-1599.
[15] Jalili-Marandi, Vahid, Zhiyin Zhou, and V. Dinavahi. "Large-Scale Transient Stability Simulation of Electrical Power Systems on Parallel GPUs." IEEE Transactions on Parallel and Distributed Systems 7.23 (2012): 1255-1266.
[16] Liu, Yunfei, and Quanyuan Jiang. "Two-stage parallel waveform relaxation method for large-scale power system transient stability simulation." IEEE Transactions on Power Systems 31.1 (2016): 153-162.
[17] Tomim, M., J.R. Martí, and L. Wang, Parallel solution of large power system networks using the Multi-Area Thévenin Equivalents (MATE) algorithm. International Journal of Electrical Power & Energy Systems, 2009. 31(9): p. 497-503.
[18] Tomim, M.A., T. De Rybel, and J.R. Martí, Extending the Multi-Area Thévenin Equivalents method for parallel solutions of bulk power systems. International Journal of Electrical Power & Energy Systems, 2013. 44(1): p. 192-201.
[19] Yusof, S.B., G.J. Rogers, and R.T.H. Alden, Slow coherency based network partitioning including load buses. IEEE Transactions on Power Systems, 1993. 8(3): p. 1375-1382.
[20] Jalili-Marandi, V. and V. Dinavahi, Instantaneous Relaxation-Based Real-Time Transient Stability Simulation. IEEE Transactions on Power Systems, 2009. 24(3): p. 1327-1336.
[21] Jalili-Marandi, Vahid, and Venkata Dinavahi. "Relaxation-based real-time transient stability simulation on distributed hardware." North American Power Symposium (NAPS), 2009. IEEE.
[22] Soykan, Gürkan, Alexander J. Flueck, and Hasan Dağ. "Parallel-in-space implementation of transient stability analysis on a Linux cluster with infiniband." North American Power Symposium (NAPS), 2012. IEEE.
[23] Abhyankar, S. and A.J. Flueck. Real-Time Power System Dynamics Simulation Using a Parallel Block-Jacobi Preconditioned Newton-GMRES Scheme. in High Performance Computing, Networking, Storage and Analysis (SCC), 2012 SC Companion:. 2012.
[24] Aristidou, P., S. Lebeau, and T.V. Cutsem, Power System Dynamic Simulations Using a Parallel Two-Level Schur-Complement Decomposition. IEEE Transactions on Power Systems, 2015. PP(99): p. 1-12.
[25] Pruvost, F., et al. Accelerated waveform relaxation methods for power systems. in Electrical and Control Engineering (ICECE), 2011 International Conference on. 2011. IEEE.
[26] Ilic'-Spong, M., M.L. Crow, and M.A. Pai, Transient Stability Simulation by Waveform Relaxation Methods. IEEE

Transactions on Power Systems, 1987. 2(4): p. 943-949.

[27] Crow, M.L. and M. Ilic, The parallel implementation of the waveform relaxation method for transient stability simulations. IEEE Transactions on Power Systems, 1990. 5(3): p. 922-932.

[28] Hogan, Emilie, et al. "Towards effective clustering techniques for the analysis of electric power grids." Proceedings of the 3rd International Workshop on High Performance Computing, Networking and Analytics for the Power Grid. ACM, 2013.

[29] Hogan, Emilie, et al. "Comparative Studies of Clustering Techniques for Real-Time Dynamic Model Reduction." arXiv preprint arXiv:1501.00943 (2015).

[30] Sullivan, B. et al. "Faster-Than-Real-Time Power System Transient Stability Simulation Using Parallel General Norton with Multiport Equivalent (PGNME)", in 2017 IEEE Power & Energy Society General Meeting. 2017. IEEE.

[31] Brakken-Thal, S., Gershgorin's theorem for estimating eigenvalues. Source:<http://buzzard.ups.edu/courses/2007spring/projects/brakkenthal-paper. pdf, 2007.

[32] Klein, D.J. and M. Randić, Resistance distance. Journal of Mathematical Chemistry, 1993. 12(1): p. 81-95.

[33] Babić, D., et al., Resistance-distance matrix: A computational algorithm and its application. International Journal of Quantum Chemistry, 2002. 90(1): p. 166-176.

[34] Ipsen, I.C. and D.J. Lee, Determinant approximations. arXiv preprint arXiv:1105.0437, 2011.

[35] Wu, J., N.N. Schulz, and W. Gao, Distributed simulation for power system analysis including shipboard systems. Electric power systems research, 2007. 77(8): p. 1124-1131.

[36] Dmitriev-Zdorov, V. Generalized coupling as a way to improve the convergence in relaxation-based solvers. in Proceedings of the conference on European design automation. 1996. IEEE Computer Society Press.

[37] Intel MPI Library 5.1. Intel: https://software.intel.com/en-us/intel-mpi-library.

[38] MPI: A Message-Passing Interface Standard Version 3.0. 2012: https://www.mpi-forum.org/docs/mpi-3.0/mpi30-report.pdf.

[39] hMETIS – Hypergragh & Circuit Partitioning, available at: http://glaros.dtc.umn.edu/gkhome/metis/hmetis/download

[40] Li, Xiaoye S. "An overview of SuperLU: Algorithms, implementation, and user interface." ACM Transactions on Mathematical Software (TOMS) 31.3 (2005): 302-325.

[41] SuperLU, http://crd-legacy.lbl.gov/~xiaoye/SuperLU/

[42] PoweR 2030: A European Grid for 3/4 Renewable Energy by 2030. 2014: http://www.greenpeace.de/files/publications/201402-power-grid-report.pdf.